\documentclass[aps,pra,superscriptaddress,twocolumn]{revtex4-2}

\usepackage{graphicx}
\usepackage[dvipsnames]{xcolor}
\usepackage{bm}
\usepackage{physics}
\usepackage{amsthm}
\usepackage{mathtools}
\usepackage{hyperref}
\usepackage{bbm}
\usepackage{subcaption}
\usepackage{amssymb}
\usepackage[T1]{fontenc}

\newcommand{\ba}{\begin{eqnarray}}\newcommand{\ea}{\end{eqnarray}}\newcommand{\ban}{\begin{eqnarray*}}\newcommand{\ean}{\end{eqnarray*}}

\begin{document}

\title{Three sins against physics by an exaggerated quantum information perspective}

\author{Valerio Scarani}
\affiliation{Centre for Quantum Technologies, National University of Singapore, 3 Science Drive 2, Singapore 117543}
\affiliation{Department of Physics, National University of Singapore, 2 Science Drive 3, Singapore 117542}

\date{\today}

\begin{abstract}
I point out three ways in which the perspective of quantum information may lead to distorted claims about physics: forgetting that light does not need to be quantised to show coherence; ignoring the generators of unitary evolutions; and approaching the discovery of nature as a fight against an adversary.
\end{abstract}

\maketitle

Reality is complex. Any new viewpoint that provides new perspectives is precious. The viewpoint of information has brought a lot to our understanding of quantum physics. The fact that it also provides us with promising technologies is correlated: we understand better when we play. But humankind cannot bear too much reality \footnote{T.S.~Eliot, Burnt Norton, Four Quartets}: we can be so excited by what we discover, that we quickly drift to believe that our viewpoint is \textit{the} viewpoint. Case in point, the quantum information viewpoint risks to hide or deform some aspects of reality. This note addresses what I think are three such situations.

\section{Thou shalt not confuse the coherence of modes and the coherence of states in optics}

The discovery of the usefulness of ``foundational'' features such as superposition and entanglement has triggered great interest in demonstrating them. Interferometers have been a tool of choice for many such studies: from the obvious one, namely demonstrating coherence through superposition, all the way to advanced ideas like comparing the three-slit interference pattern with the two-slit ones, or the simulation of beyond-quantum processes like ``quantum switches''. When interferometry is implemented with matter waves, the coherence is indeed the one of quantum theory. Things are different if one uses optical interferometry. 

In quantum optics, two Hilbert spaces are at play, each with its own coherence \cite{RevModes}. One is the Hilbert space of optical modes, with the superposition of fields implied by the linearity of Maxwell's equations. This is common to classical and quantum optics, and is well known to show interferences since Young used them in 1805 to prove that light is a wave. In the quantum theory of light, one associates a Hilbert space to each mode, to describe the quantum states of the mode. The collection of these spaces, called the Fock space, is our second Hilbert space. 
 
Thus, if one wants to claim fundamental quantum tests in optical setups, one has to prove that the observation requires the Fock space and cannot be explained by the classical coherence alone. In other words, one should ask: am I seeing a signature of the quantum state? The answer is not trivial and may be positive. It is textbook knowledge that $g^{(1)}$ informs you about the mode and not the state, while $g^{(2)}$ may carry information on the state (although the Hanbury Brown-Twiss experiment is explainable with classical optics). Another known positive example is boson sampling \cite{aa13}, where one measures detailed counting statistics, which are a signature of the quantum state with no classical analog. However, \textit{if one only estimates intensities, even reconstructed from photon counting, only the the superposition of modes matters}. For instance, if detection probabilities in a linear optical interferometer were to differ from the square of the amplitudes, this would not signal a deviation from the Born rule: it would signal that the electromagnetic energy is not the square of the fields.

Using the word ``photon'' does not save the day. Since this claim has a long history of stirring controversy \cite{lamb}, let me expand. Feeding a linear interferometer with a single-photon source will test the coherence of the mode created by that source: it is an important characterization of that device, but does not involve any coherence of the state. As to whether ``single-photon detection'' is a purely quantum process, it may depend on the principle underlying the detection; the vast majority of detectors are based on the photoelectric effect, in which case the answer is negative, despite Einstein's intuition. Indeed, operating the detector ``in the single-photon regime'' means that an amount of energy $E\lesssim h\nu$ is sufficient to trigger the amplification of the electric signal to recordable levels. This threshold behavior does not depend on the fact that $h\nu$ is the smallest excitation of the field (although you know that it is pointless to amplify even more, trying to detect excitations at $E\ll h\nu$). And another argument: in the chapter on time-dependent perturbation theory of any undergraduate quantum mechanics course, one finds that the photoelectric effect requires only the quantisation of the detector, not that of the field.

So, the first warning of this paper is: be careful in claiming ``quantum'' when your setup is optics (I focused on linear optics and interferences; also non-linear optics has classical aspects). Before moving on, I confess to having committed this sin early in my career \cite{ss98}.

\section{Thou shalt not forget the Hamiltonian that generates your unitary}

Quantum information statements and protocols are platform-independent. Every two-level system is a qubit; entanglement does not depend on whether the degrees of freedom belong to two photons, to two atoms, to one photon and an ion... Quantum circuits are usually described in terms of \textit{unitary gates}, leaving it to the hardware platforms the task of implementing them; entanglement is invariant under \textit{local unitaries}... Unitaries are everywhere.

Still, these unitaries must be realized by the action of some Hamiltonian $H$ for a time $\tau$. In the simplest case, if the Hamiltonian is time-independent (besides being switched on and off), we have $U=e^{-iH\tau/\hbar}$. The risk of the quantum information perspective is to develop a ``gate mentality'': thinking only in terms of unitaries, forgetting the Hamiltonians. 

Inspired by classical logical circuits, the practice of working with a restricted set of unitaries (``gates'') has shaped the field from the start, as the basis for the circuit model of quantum computing. Meanwhile, restricting operations is also a key step in many \textit{resource theories}. In these theories, inspired by the way entanglement theory was set up \cite{horo}, one defines a resource (e.g.~entanglement), the states that don't have it (e.g.~separable) and the operations that don't generate it (e.g.~local operations and classical communication). These approaches proved overwhelmingly fruitful. On the wings of enthusiasm, one risks to forget that \textit{a restriction of the set of unitaries describes a physical constraint on the available resources only if the restriction is actually on the Hamiltonian}. Among the most famous examples, the restriction is on the Hamiltonians for the local operations of entanglement theory, and for Gaussian operations on continuous variables \cite{gauss}; it is not for Clifford operations on discrete variables \cite{GK}. 

Take the case of $U=\sigma_x$. The simplest way to generate it is to use $H=g\sigma_x$ for a time $\tau$ given by $g\tau/\hbar=\pi/2$. This means that whoever can implement $U(\frac{\pi}{2})=\sigma_x$ (up to a phase) has the resources to implement $U(\theta)=e^{-i\theta\sigma_x}$ for any value of $\theta$; certainly for smaller values, if one is afraid of decoherence for longer times. These shorter-time unitaries contain magic and generate a lot of coherence in the computational basis. By the same argument, whoever can implement all Clifford gates can also generate a lot of magic and coherence. There is no lab that can implement \textit{only} the Clifford gates. %Starting with the Gottesman-Knill theorem, most results neither assume nor need such a physical assumption. 

%The second situation has to do with energy. Suppose that a system with bare Hamiltonian $H_0$ undergoes a unitary gate: for a time $\tau$, $H\neq H_0$ as there is an interaction term; since the interaction is switched on and off, the Hamiltonian is time-dependent, and energy may not be conserved even in classical physics. The case when it // that the bare energy $\langle H_0\rangle$ is conserved through $U$. In other words, the condition $[U,H_0]=0$ is not ``conservation of energy'': its name in physics is ``elastic scattering''. By that semantic confusion, an audience trained in the non-physics disciplines of quantum information may believe that $[U,H_0]=0$ is a mathematical assumption but a physical necessity. The reality is that $[U,H_0]=0$ is violated by almost every driven unitary a physicist can think of: for instance, a $\pi$-pulse switches the ground and the excited state of a qubit (an implementation of $\sigma_x$, by the way).

So we have come to the second warning of this paper: if your work is phrased only in terms of unitaries, check if your narrative survives once you look at these unitaries as ``Hamiltonians applied for some time''. I don't recall having much to confess here, but for having occasionally given in to the practice of calling ``energy conservation'' a condition that should be properly called ``elastic scattering''.

\section{Thou shalt not confuse Mother Nature with Eve}

The ontology of quantum mechanics has been the subject of heated discussions since its beginning, and there does not seem to be an end to it in sight. In the study of Bell nonlocality \cite{scarani}, the quantum information perspective brought a very welcome gust of fresh air through the phrasing in terms of \textit{nonlocal games} \footnote{To the best of my knowledge, this framework was first proposed in 2004: R.~Cleve, P.~Hoyer, B.~Toner, J.~Watrous, in 19th
IEEE Conference on Computational Complexity, p. 236.}. The physicists become players, who win the game if they can produce the correlations predicted by quantum theory; which they can do if they share quantum systems of sufficiently high quality. Actually, depending on the papers, the players may not be the physicists but the quantum systems themselves. For clarity, here Alice and Bob are physicists, their quantum systems will be called their ``particles''.

The game perspective is adversarial: if the players were unwittingly allowed to cheat, they would cheat and claim a win without the quantum resources. The fact that Bell's theorem holds in this setting gives a clear feeling of its power: not many physics results would stand such a scrutiny, and based on such high-level assumptions. For purposes of certification and security (randomness, cryptography), this perspective is almost unavoidable. But it is not compulsory for ontology, at least insofar as one believes that Mother Nature is not adversarial.

Let me expand this observation in one possible direction. The game perspective assumes the cryptographic rule ``no security by obscurity'': the players know which game they are going to play, the only thing they should not know are the inputs they will receive in a specific round. The physicists in the lab also know what experiment they want to perform. But it is natural (or at least, neither unnatural nor absurd) to believe that \textit{the particles do not know which experiment they are going to undergo}. For instance, only at detection is the particle ``forced'' to give an outcome, based on all that it has gone through until then. Thus, only at detection would it make sense to inform the other particle of all that happened, if a yet-unknown mechanism for sending such a signal was available. So, to understand ``how nature does it'' in a non-adversarial perspective, the locality loophole seems to be far less stringent than in an adversarial setting -- if it even exists \footnote{The usual approach to the locality loophole was meant to convince people who may believe that unknown influences carry information around at the speed of light, but not faster. The idea is that spacelike separation should then convince them -- but spacelike separation of which events? People who believe in unknown forms of communication may also believe that the microscopic events leading to the choice of the input had started well before the random number generator sent out a readable signal: that form of the locality loophole cannot be closed.}.

In conclusion, my third and last warning: when trying to understand Mother Nature, you may legitimately deviate from the game perspective and its operational approaches to independence. Don't let quantum information people bully you into using definitions that are meant for a cheating Eve \footnote{Some may argue that adopting an adversarial view is ``more general''. If that were true, our prior should be that most experiments in quantum information are fake (since the results were easily predictable by the theory), leaving to our colleagues from the lab the burden to convince us that they actually performed the experiment. I for one would not want to work in such a toxic climate. And if we trust our colleagues, why should we not trust nature, if our goal is to describe it? Even Descartes had to ground his systematic doubts on a denial of ultimate malignity.}. I may have been one such bully in some past discussions; at least, I remember having believed for a long time that those definitions were the only correct ones.

\section*{Acknowledgments}

I acknowledge support by the National Research Foundation, Singapore through the National Quantum Office, hosted in A*STAR, under its Centre for Quantum Technologies Funding Initiative (S24Q2d0009).

\bibliography{references}

\end{document}